\documentclass[intlimits,twoside,a4paper]{article}

\usepackage[T2A,T1]{fontenc} %%%% already in cmpj3.sty
\usepackage[cp1251]{inputenc}
\usepackage{amsmath,amssymb}

\usepackage[eqsecnum]{cmpj3}
\issue{2023}{26}{4}{43602}
\doinumber{10.5488/CMP.26.43602}

\title[Organizing MgO and ZnO NPs in 6CHBT LC media]%
{Formation of  nano and micro scale hierarchical structures in MgO and ZnO quantum dots doped LC media: The role of competitive forces}%
	
\author[A. K. Singh, S. P. Singh]{A. K. Singh\orcid{0000-0002-8075-806X},
	S. P. Singh\orcid{0000-0002-5002-8787}\thanks{Corresponding author: \email{singh.satyapal@hotmail.com}.}}
\address{
	Department of Physics and Material Science, Madan Mohan Malaviya University of Technology, Gorakhpur-273010, Uttar Pradesh, India
}


\date{Received April 17, 2023, in final form July 03, 2023}

\begin{document}

\maketitle

\begin{abstract}
In this paper, we have studied the effect of doping of ZnO and MgO nanoparticles (NPs) in 4-(trans-4-n-hexylcyclo\-hexyl) isothiocyanatobenzoate. A thorough comparison of dielectric properties, optoelectronic properties, and calorimetric phase transition properties has been done for MgO and ZnO NP doped LC. We prepare their homogenous mixture of MgO and ZnO NPs in toluene and transfer into cells made of glass and Indium Tin-Oxide (ITO) coated glass. The observed microstructures in the hybrid system can be classified into three main categories: grain like structures formed by aggregation of smaller size MgO nanoparticles while liquid crystal molecules anchor over the surfaces of nanoparticles, the grtu grain-like structures further integrate to form inorganic polymeric type of honeycomb-like mesostructures in presence of glass surface, and flower-like clusters of MgO nanoparticles on ITO surface. The smaller size nanoparticles can maintain the energy balance by allowing the anchoring of liquid crystal molecules over their surfaces whereas the larger size nanoparticles cannot compromise or maintain the energy balance with the liquid crystal molecules and are separated out to nucleate and form bigger size nanoaggregate or clusters. The energy preference of the substrate and nanoparticle’s surface to liquid crystal molecules plays an important role in the formation of different types of hierarchical nano- and microstructures. We account the reasons for the formation of nano and micro scale hierarchical structures on the basis of the competition between the forces: NP--NP, LC--LC, NP--LC, Glass/ITO-NP, and Glass/ITO-LC interactions. We observed a considerable change in the dielectric properties, transition temperature, bandgap, and other parameters of LC molecules when MgO NPs are doped, but a minor change occurs when ZnO NPs are doped in LC. Optical microscopy, FTIR, Raman, IR, HR--XRD and FESEM-EDX characterization data confirm and validate our guiding conceptions.
%
%
%\keywords Up to six keywords (\href{https://physh.aps.org/browse}{Physics Subject Headings})
\printkeywords
%
%\pacs Up to six PACS numbers (optional)
\end{abstract}

\section{Introduction}

\begin{figure}[htb]
	\centerline{\includegraphics[scale=0.8]{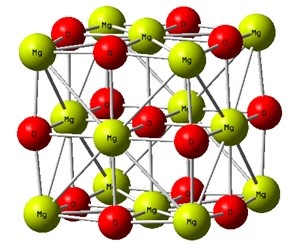}}
	\caption{(Colour online) Crystal structure of MgO nanoparticle obtained using HF method with basis set 6-31G using Gaussian 9 software.} \label{fig-smp1}
\end{figure}

Many metal oxide nanostructures, including Fe$_{3}$O$_{4}$, ZnO, TiO$_{2}$, MgO, and CoFe$_{2}$O$_{4}$, are widely used in many applications. Semiconductors composed of elements of II and VI groups, e.g., ZnO, MgO, CdS and PbS, etc., are a unique class of substances having characteristics intermediate bandgaps between insulators and conductors. Modern material science considers ZnO and MgO to be significant II--VI semiconductors with a number of distinct properties, which make ZnO and MgO versatile materials suitable for a variety of novel applications, including light-emitting diodes, ultraviolet (UV) lasers, solar cells, solar water splitting cells, biosensors, and piezoelectronics~\cite{RS1}. ZnO and MgO nanoparticles (NPs) possess wide bandgaps in the range of 4.0--5.0~eV showing good transparency in the visible range. Doping in metal oxide NPs help to engineer their bandgap. MgO is a promising inorganic material that crystallizes in NaCl type crystal structure shown in figure~\ref{fig-smp1}. It is widely used for various applications such as sensors, antibacterial agents, optical coatings, water treatment, catalysis, adsorption, etc., as well as fuel additives, among others~\cite{RS2,RS3,RS4}. This is primarily due to its high surface reactivity, broad band-gap, and chemical and thermal stabilities. Researchers have concentrated on the synthesis of MgO nanoparticles and their nanocomposites due to its vast applications in many advanced and sophisticated technologies. Additionally, nanostructured MgO is used in a variety of places, such as making ceramics, and electronic items as well as catalysts. Non-toxic MgO nanoparticles are utilized in coating materials, crucibles, and as additives in flame-retardant and refractory materials~\cite{RS5}. Particle size of the MgO NPs and its morphologies can be controlled by adjusting variables including pH, ionic strength, and calcination temperature using chemical methods. Sol-gel, hydrothermal, spray pyrolysis, combustion, microwave, and co-precipitation processes are among the various preparation techniques used to produce MgO nanoparticles; each has its own advantages and disadvantages~\cite{RS6,RS7,RS8,RS10}.

The US Food and Drug Administration (FDA) has approved the use of ZnO in cosmetic products, which essentially makes it one of the important factors in the growth of the ZnO market~\cite{RS11}. ZnO is also recognized as a safe biocompatible material. Being physiologically active and capable of forming reactive oxygen species by releasing Zn$^{2+}$ ions, ZnO nanostructures are always accompanied by toxicological tests to determine their biocompatibility, although many studies have demonstrated that ZnO NPs possess antibacterial and antifungal properties. This is in addition to their properties of enhancing cell growth, differentiation, and proliferation as well as tissue regeneration, boosting angiogenesis and osteointegration processes. Moreover, ZnO nanostructures show a selectivity for particular cell lines, making them potential candidate for the elimination of cancer cells~\cite{RS12}.

Due to its special physical, optical, and nonlinear optical dielectric properties, liquid crystals (LCs) are used in a wide range of modern applications such as opto-electronic devices, display technologies, bio photonics, and biosciences. The optical and dielectric anisotropy of liquid crystal display a key role and is exploited in making optical and electro-optical systems. Liquid crystals have exceptional physical and electro-optical qualities that place them off edge. These qualities include quick reaction, low threshold voltage, high resolution, low weight, low power consumption, and better image quality. The response of the liquid crystals to the external electric field is governed by their dielectric properties. On the other hand, the LC molecules can reorient when an external field (electrical or magnetic) is applied. Dispersion of nanoparticles in liquid crystals for the purpose of altering, enhancing, and controlling the optical and dielectric properties has gained a lot of attention in recent years. Creating a stable dispersion of NPs to enhance the optical and electro-optical responses of the LC molecules is one of the numerous possible new and emerging areas of research. Numerous researchers have reported that the addition of NPs changes the values of the dielectric permittivity and dielectric anisotropy of LCs. The presence of zinc ferrite (ZnFe$_{2}$O$_{4}$) has a significant impact on the dielectric and electro-optical properties of the host 7CB nematic liquid crystal~\cite{RS13,RS14,RS15}.

The nanocomposites of liquid crystal with MgO and ZnO nanoparticles have a number of exceptional properties that can support its use in a variety of fields. Even at high temperatures, MgO and ZnO nanoparticles have exceptional thermal stability. They improve the thermal stability of the composite material when added into a liquid crystal matrix. Due to its property, it can be used in applications where effective thermal management is essential, such as heat sinks, thermal interface materials, and electronic devices. The mechanical strength and stiffness of the nanocomposite system composed of LC and QDs greatly increased. The matrix is strengthened by the nanoparticles, improving structural integrity and deformation resistance. The development of robust and long-lasting materials for structural applications, such as lightweight construction materials, automobile parts, and aircraft components, benefits greatly from this property. The dielectric characteristics of the liquid crystal nanocomposite can be enhanced by MgO nanoparticles. They raise the material's effective permittivity, which enhances capacitive behaviour and allows for greater charge storage. The development of high-performance dielectric materials for energy storage technologies such as capacitors and supercapacitors is enhanced due this property. The liquid crystal matrix can display special optical features when MgO nanoparticles are incorporated. They have an impact on the composite material's refractive index and light scattering properties. The creation of advanced optical devices, such as waveguides, optical switches, and tunable photonic crystals, are possible using this property. MgO nanoparticles can improve the electrical conductivity of a liquid crystal nanocomposite. They may improve the material's electrical characteristics by facilitating the charge flow through it. This property is useful in printed electronics, flexible circuits, and conductive coatings. The MgO, ZnO NPs-LC nanocomposite can be used for environmental sensing applications such as environmental monitoring, pollution detection, and chemical sensing. These exceptional properties emphasize the applications of the nanocomposite of MgO and ZnO nanoparticles and liquid crystals in a variety of domains, including electronics, optics, energy storage, sensing, and biological applications~\cite{RSS16,RSS17,RSS18}. Continuous research and development in this area will expand the possibilities and drive new applications.

Creation of a new kind of stimuli-responsive functional composite material made of nanoparticle dispersed liquid crystals is possible and may get possible commercial applications in near future. By carefully controlling the positional and orientational orders of NPs in LC media, the interparticle in\-te\-rac\-tions because the topological flaws in the LC may be programmed. The interactions of NPs with LC molecules and coupling with the director field result in positional and orientational changes in LC media. The organization of LC molecule is strongly influenced by anchoring of LC mesogens at the surface of NPs. The excess surface free energy of NPs, chemical incompatibility with the LC medium and the energetic variations caused by the director field distortion affects the NP dispersion and its stability. Unfavourable surface contacts of NPs with LC molecules cause NPs to phase separate if the director field distortion of the LC host molecules expel them to reduce the total elastic energy. The entropic contributions also balance these interactions. The anchoring conditions must be matched with the symmetry of NPs in order to reduce or completely eliminate the director field distortions. With little or no director field distortion, homeotropic and planar anchoring at flat surfaces and planar anchoring perpendicular to a cylindrical or rod-like NP can be achieved. By altering the NP surface chemistry with the right ligands, it is possible to compensate the excess energy of chemical incompatibility of NPs~\cite{RS16,RS17,RS18,RS19}.

Dipole-dipole interactions, London dispersion forces (commonly referred to as van der Waals forces), and hydrogen bonds are the three main types of intermolecular non-bonding interactions, which play a vital role in molecular scale reorganization especially in the case of macromolecules. Intermolecular interactions hold molecules with one another in liquids and can hold polyatomic ions together. The intramolecular forces are generally of a bonding nature and hold the atoms within molecules together. Intermolecular interactions are weaker than intramolecular interactions. Intermolecular interactions vary as the phase changes from the solid to liquid or from the liquid to the gas phase, whereas the intramolecular interactions are unaffected by the phase change. The electrostatic interactions between positive and negative ends of a polar molecule give rise to dipole-dipole interactions and the molecule has a permanent dipole moment. The strength of the dipole-dipole interaction varies as ${1}/{r^{3}}$, where $r$ is the distance between the two dipoles. The short-lived fluctuations in the electronic charge distribution that leads to the creation of as instantaneous dipole moment in polar or nonpolar molecules causes the formation of an induced dipole moment in nearby molecules temporarily, which varies as ${1}/{r^{6}}$.

The microstructures formed using NPs have novel applications in various fields. The possible compositions and different mechanisms for the formation of hollow nano- and microstructures are presented and also highlighted its various applications~\cite{RS20,RS21,RS22}. The hybrid nano- and microstructures show superconducting properties. Aromatic material lignin which is spherical nanoparticle also forms microstructures~\cite{RS23}. These microstructures have novel biomedical applications, drug vehicles for pharmaceutical ingredients and in treatment of numerous diseases including cancers. Researchers have summarized the applications of mixture of E7 nematic LCs and polydimethylsiloxane (PDMS) polymeric material~\cite{RS24}. They created an electrically controlled liquid crystal and polymer microstructures and have suggested its potential sensing applications. Significant studies have been done on the structural, optical, and antibacterial characteristics of MgO nanoparticles~\cite{RS25,RS26,RS27,RS28}. Dispersion of ZnO nanoparticles in different concentrations in LC compounds as p-n-propoxy benzoic acid (3oba), p-n-propyl benzoic acid (3ba) and its mixtures has been already investigated~\cite{RS29}. The authors have observed that there is no significant effect on the ordering and textures of LC molecules. We have also tried to observe the effect of doping of ZnO NPs in 4-(trans-4-n-hexylcyclohexyl)isothiocyanatobenzoate (6CHBT) LC media. The selection of 4-(trans-4-n-hexylcyclohexyl)isothiocyanatobenzoate (6CHBT) as a liquid crystal rather than the more often used 4-cyano-$4'$-pentylbiphenyl (5CB) is due to the special properties and features that make 6CHBT more advantageous. 6CHBT possesses a stable smectic phase across a wide temperature range, including room temperature. This stability is critical for applications requiring a well-defined smectic phase. By contrast, the smectic phase of 5CB has a narrow temperature range. When compared to 5CB, 6CHBT typically has faster electro-optical response times.

\begin{figure}[htb]
	\centerline{\includegraphics[scale=0.7]{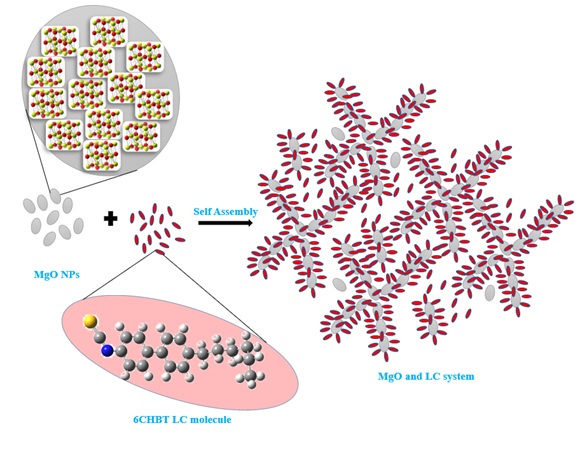}}
	\caption{(Colour online) Illustration of the self-assembly of nanoparticles.} \label{fig-smp2}
\end{figure}

Doping MgO and ZnO NPs into 6CHBT [4-(trans-4-n-hexylcyclohexyl)isothiocyanatobenzoate] LCs can alter their properties. Some of the characteristics that can be altered through nanoparticle doping are as follows: doping NPs can change the electro-optical response of 6CHBT LCs. The inclusion of MgO and ZnO NPs can affect the reaction time and threshold voltage of liquid crystals, resulting in quicker switching rates and lower operating voltages. The inclusion of MgO and ZnO nanoparticles can alter the refractive index, birefringence, and light scattering properties of liquid crystals. Due to their strong thermal conductivity and stability, MgO and ZnO nanoparticles can assist to prevent liquid crystal breakdown at high temperatures. MgO and ZnO NPs can improve the mechanical strength and stiffness of 6CHBT liquid crystals. The addition of MgO and ZnO nanoparticles into the liquid crystal matrix can increase charge transport characteristics, resulting in a better electrical conductivity. MgO and ZnO nanoparticles can affect the surface contact forces, causing changes in the alignment and orientation of liquid crystal molecules. By doping MgO and ZnO nanoparticles in 6CHBT liquid crystals, these properties can be modified and the performance of the liquid crystal system can be tuned for specific applications such as displays, sensors, and other electro-optical and opto-electronic devices. Many researchers have studied the effect of MgO doping in 6CHBT LC media and its effect on optical and dielectric properties of LC molecules. However, an explanation for the formation of mesostructures on the basis of molecular interactions has not been discussed much. We have investigated the formation of nano-and microstructures in MgO and ZnO doped in 6CHBT LC medium and have accounted reasons for their formation on the basis of competitive forces yielding minimum energy of the hybrid system. The interactions can be broadly considered as NP--NP, LC--LC, NP--LC, Glass/ITO--NP, and Glass/ITO--LC molecules. We have successfully explained the effect on the optoelectronic properties of the LC molecules on the basis of structural reorganization after doping NPs in LC system. We have investigated shifting in some of the key features, as shown in the table~\ref{tbl-smp3} below. The molecular scale reorganization of MgO NPs doped in LC media is illustrated in figure~\ref{fig-smp2} and is very relevant and useful for explanation on the basis of molecular interactions as discussed above. The figure only presents the schematic representation for the formation of hierarchical nano-micro structures after self-assembly of NPs in presence of LC media.

\begin{table}[htb]
	\caption{The observed variations in the properties of the MgO, ZnO, and LC matrix. }
	\label{tbl-smp3}
%	\vspace{2ex}
	\begin{center}
		\renewcommand{\arraystretch}{0}
		\begin{tabular}{|c|c|c|c|}
			\hline
			Properties& Pure LC &MgO doped LC&ZnO doped LC\strut\\
			\hline
			\rule{0pt}{2pt}&&&\\
			\hline
			Bandgap, eV & 3.49 &4.26 & 4.28\strut\\
			\cline{1-4}
			Transition temperature, °C& 43.23   &42.82 & 43.09  \strut\\
			\cline{1-4}
		Dielectric permittivity $\epsilon_{\perp}$ (at 50 Hz)& 6.12 &5.80&6.00\strut\\
			\cline{1-4}
			Dielectric permittivity $\epsilon_{\parallel}$ (at 50 Hz)& 41.98&22.99&35.00\strut\\
			\cline{1-4}
			\hline
		\end{tabular}
		\renewcommand{\arraystretch}{1}
	\end{center}
\end{table}

\section{Experimental details}

\subsection{Materials used}

Mg(CH$_{3}$COO)$_{2}$ (magnesium acetate tetrahydrate, purity 99.5\%), H$_2$C$_2$O$_{4.2}$H$_2$O (oxalic acid dihydrate, purity $>98 $\%), and CH$_3$OH (methanol, purity 99.9\%) are used to synthesize magnesium-oxalate precursor, which is further converted into MgO nanoparticles. All chemicals are purchased from Avantor Performance Materials India Limited and SDFCL. For the synthesis of ZnO NPs we have used reagent grade zinc acetate dihydrate [Zn(CH$_{3}$COO)$_{2.2}$H$_{2}$O, 98\% purity], sodium hydroxide pellets (NaOH, 97.0\%), ethanol (CH$_{2}$COOH, 99.9\%) and de-ionized water. The materials were purchased from Sigma-Aldrich. Zn(CH$_{3}$COO)$_{2.2}$H$_{2}$O is used as a precursor, ethanol is used as a reagent and the deionized water is used a solvent medium. 6CHBT liquid crystal (99\% pure) was bought from Sigma Aldrich and is used without further purification. Other items include 60 \textmu{}m Mylar strips used as spacers, acetone, etc. Only analytical research grade chemicals were used in performing the experiment. 

\subsection{Synthesis of MgO and ZnO quantum dots}
\begin{figure}[htb]
	\centerline{\includegraphics[scale=0.4]{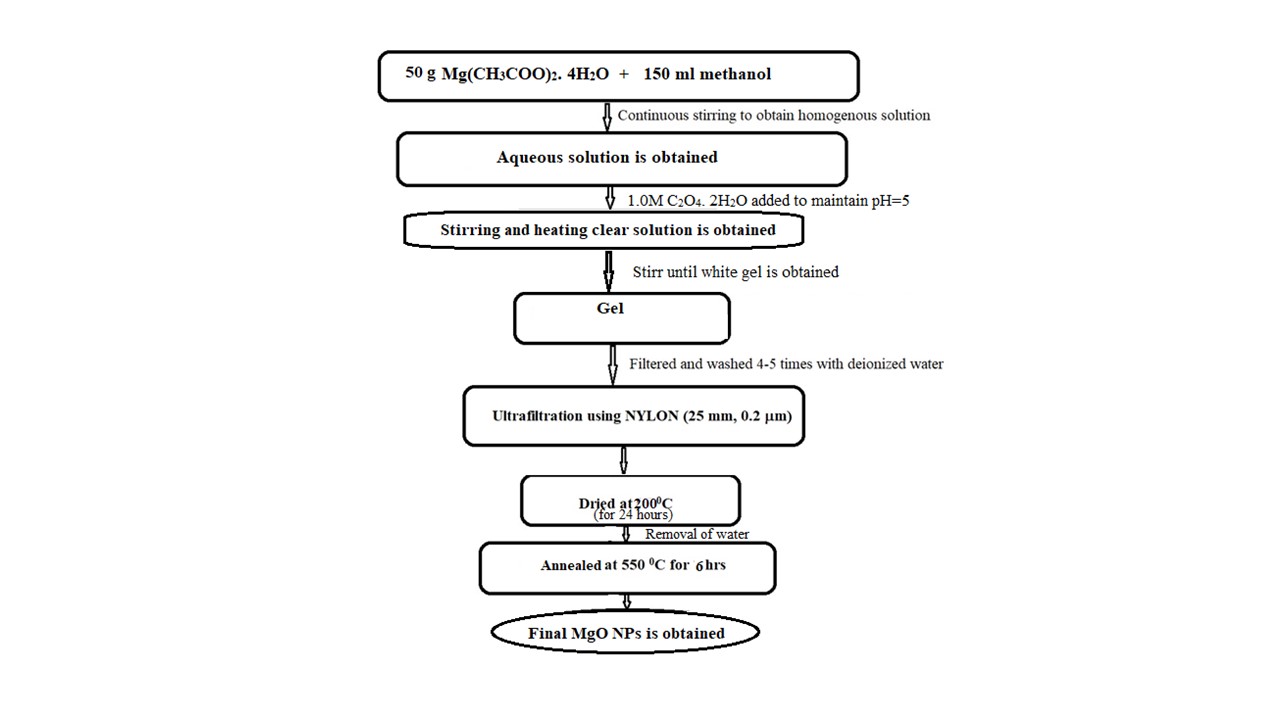}}
	\caption{Shows different steps involved in the synthesis of MgO QDs.} \label{fig-smp3}
\end{figure}

The flowchart for the synthesis of MgO quantum dots (QDs) is shown in figure~\ref{fig-smp3}. We followed the procedure using Mg(CH$_{3}$COO)$_{2.4}$H$_{2}$O and C$_{2}$O$_{4.2}$H$_{2}$O materials. We first dissolved these materials in methanol. About 50~g of Mg(CH$_{3}$COO)$_{2.4}$H$_{2}$O was mixed in 150 ml methanol by continuous stirring for 30 minutes using magnetic stirrer at constant rpm of 800 and at room temperature. Then, we stir and heat the solution until a clear solution is obtained. pH~$=5$ is maintained via adding 1.0~mol oxalic acid solution in the resulting mixture. Then, we continuously stir the mixture until a white gel is obtained. To complete the gelation process, the gel is kept at room temperature for overnight. The gel is filtered using a filter paper and is separated from the solution. This separated gel is centrifuged at 4000~rpm and washed with ethanol at a speed of 4000~rpm for 5~min in centrifuge machine. Further, filtering is done by using mesh of ultra-fine filter made of nylon 25~mm 0.2~\textmu{}m (AXIVA). The finally obtained powder is heated at 200°C for 24 hours to remove the trapped acetate and water. Furthermore, the obtained dried product is crushed to produce fine powder using mortar and pestle. The finally formed complex was annealed at 550°C for 6 hours in a Muffle furnace. We finally obtain crystalline fine powder of MgO QDs.

In a similar way, we have synthesized ZnO QDs. We take 6 gm of zinc acetate dihydrate and dissolve in 45 ml of distilled water and 120 ml ethanol (say solution A). A separate solution is obtained by dissolving 24 gm sodium hydroxide into 30 ml of distilled water and 75 ml ethanol (say solution B). We then place each of the solutions separately at magnetic stirrer at 500 rpm at room temperature for an hour to obtain microscopically homogenous solutions. To obtain a homogeneously mixed solution, the solution B is added drop-wise into the solution A with constant stirring speed of 800 rpm for 2.25 hours at 50°C. After cooling the mixed solution at room temperature, we then centrifuge the mixture at 4000~rpm and wash the obtained QDs with ethanol at the 4000 rpm for 5 min using centrifuge machine. After washing ultra-filtration is done using mesh of ultra-fine filter made of nylon 25~mm 0.2~\textmu{}m (AXIVA). We then obtain nano-powder and dry it in a laboratory oven for two hours at 60~°C. We finally obtain the desired ZnO QDs.

\subsection{Preparation of hybrid system of LC and nanoparticles}
\begin{figure}[htb]
	\centerline{\includegraphics[scale=0.4]{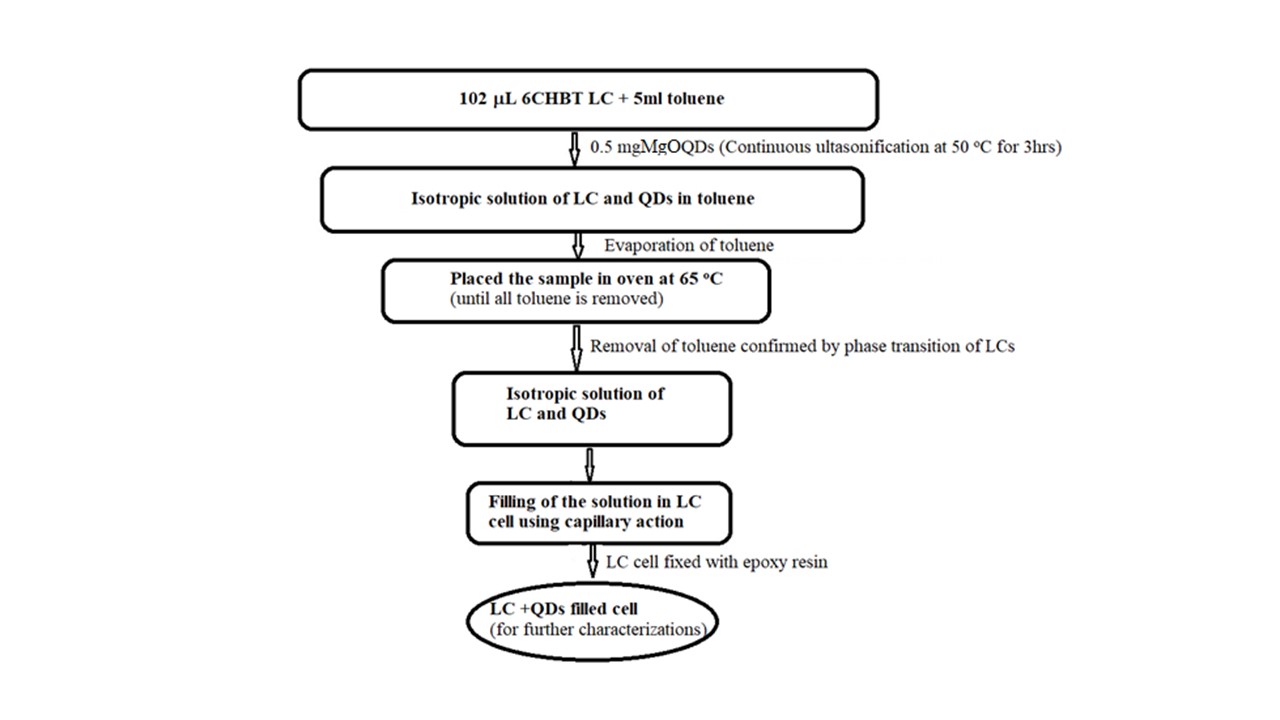}}
	\caption{Shows different steps in the formation of a hybrid system of 6CHBT liquid crystal and MgO nanoparticles hybrid system.} \label{fig-smp4}
\end{figure}

The homogenous and isotropic mixture of liquid crystals and NPs was prepared by dissolving the two components in toluene. We used MgO and ZnO both in 0.5 weight percent concentrations in liquid crystals. Then we slowly evaporate the solvent, by drying it in vacuum. Figure~\ref{fig-smp4} shows a flowchart for the production of an LC and MgO nanodot hybrid system. To generate a homogeneous solution, we bath sonicated 0.5 mg MgO NPs, 102~\textmu{}l 6CHBT LC in 5 ml toluene solution for 3 hours at 50°C. We obtain a clear solution, and place it in an oven at 60°C to evaporate the remaining toluene. The phase of the LC remains unchanged during the creation of the isotropic solution. The sample was heated to 65°C to thoroughly evaporate the undesired solvents used to prepare the homogeneous solution of LC and MgO dots, in order to transferr it into the cells. It can be easily demonstrated that 6CHBT becomes isotropic above 45°C, and returns to the nematic phase, when cooled below 45°C. By sandwiching Mylar strips (spacers) between a substrate and a cover, 60~\textmu{}m thick cells were created. We do not distinguish between substrate and cover, in case we use the same materials for both. Planar texture is formed on the substrate by rubbing it with a cotton cloth. Finally, we use capillaries to fill the cell with a homogeneous mixture. Similarly we have done with ZnO NPs and the LC system for synthesizing the hybrid system. We proceed for sample characterization after sealing the cells from all sides with epoxy resin.

\section{Result and discussion}

\subsection{Fourier transform infrared spectroscopy (FTIR)}
\begin{figure}[htb]
	\centerline{\includegraphics[scale=0.6]{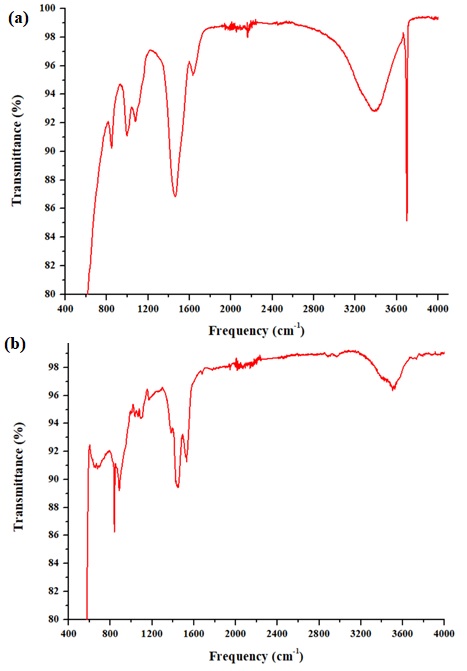}}
	\caption{(Colour online) FTIR spectra of synthesized (a) MgO and (b) ZnO quantum dots.} \label{fig-smp5}
\end{figure}

FTIR spectra are observed using Shimadazu IR Sprirt FTIR-8000 apparatus. FTIR spectrum of MgO and ZnO QDs synthesized using sol-gel method is shown in figure~\ref{fig-smp5}. Dominant peaks are observed in transmission mode in the wavelength range 400~cm$^{-1}$ to 4000~cm$^{-1}$. Figure~\ref{fig-smp5}(a) shows the FTIR spectrum of MgO nanoparticles. The peaks at the 652~cm$^{-1}$ and 836~cm$^{-1}$ indicate for the presence of Mg--O--Mg bonds~\cite{RS25}. The bending vibration of the surface hydroxyl group is represented by a unique band at 1002~cm$^{-1}$ and 1074~cm$^{-1}$. Due to the O--H stretching vibration of the water molecule, a broad band of 3395~cm$^{-1}$ is detected. It indicates the presence of the adsorbed water molecules due to exposure to the atmospheric conditions. NPs surface is reactive and can easily absorb water and carbon dioxide molecules present in the ambient environment. A very weak band at 2165~cm$^{-1}$ shows the adsorption of CO$_{2}$ molecule. Sharp peak at 1460~cm$^{-1}$ corresponds to the presence of CO$_{3}^{-}$. Thus, our FTIR analysis clearly confirms that the sol-gel method used in our experiment successfully yields MgO NPs. The presence of water molecules is also confirmed due to absorption by MgO NPs.

Figure~\ref{fig-smp5}(b) shows the FTIR spectrum of ZnO nanoparticles. A broad peak around 3441~cm$^{-1}$ and 3105~cm$^{-1}$ is due to the presence of OH stretching vibrations. The peaks around 1611~cm$^{-1}$  and 1555~cm$^{-1}$ are due to the vibrations of C=O bond. The bond arising at 1440~cm$^{-1}$ and 1392~cm$^{-1}$ is due to the presence of -C-H bending vibrations. The band at 624~cm$^{-1}$ and 620~cm$^{-1}$ is due to the O--H bending vibrations. Peaks at 568~cm$^{-1}$ and 465~cm$^{-1}$ correspond to the stretching vibrations of Zn--O bonds.

\begin{figure}[htb]
	\centerline{\includegraphics[scale=0.6]{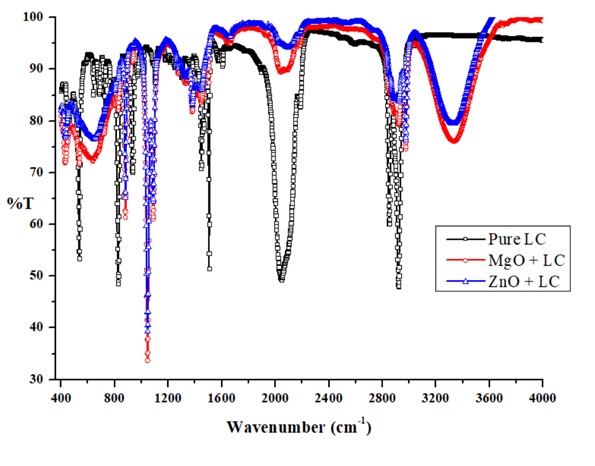}}
	\caption{(Colour online) FTIR spectra of pure LC, MgO, and ZnO doped LC matrix.} \label{fig-smp6}
\end{figure}

Figure~\ref{fig-smp6} displays the FTIR spectra of pure 6CHBT LC, MgO and ZnO NPs doped 6CHBT hybrid systems in transmission mode. The spectra are obtained in the range 400~cm$^{-1}$ to 4000~cm$^{-1}$. In the case of a pure LC system, the symmetric and asymmetric stretching modes of the carbon atoms comprising the chain create prominent peaks in the region of 2850~cm$^{-1}$ to 3000~cm$^{-1}$. The prominent peak at 2929~cm$^{-1}$ corresponds to asymmetric C--H bond stretching in the CH$_{2}$ groups of n-hexane and cyclohexane. The stretching of the C=N bond in the NCS group is detected at 2050~cm$^{-1}$. Due to the presence of multiple interactions involving the NCS group, this band becomes wider. Another peak at 1505~cm$^{-1}$ is associated with C=C in aromatic rings.    The 1456~cm$^{-1}$ frequency is connected with scissor-like deformations in the -CH$_{3}$ and -CH$_{2}$ groups. The relatively small band at 932~cm$^{-1}$ is related with C=S bond elongation. The peak at 823~cm$^{-1}$ arises due to the aromatic hydrogen with out-of-plane bending vibrations. The band at 536~cm$^{-1}$ corresponds to the -NCS group with in-plane bending vibrations or out-of-plane bending vibrations. After doping this peak shift to 535~cm$^{-1}$ for the case of MgO doping and 534~cm$^{-1}$ for the case of ZnO doping. The spectra of the MgO and ZnO doped LC system confirm the formation of a composite system via non-bonding interactions. After doping, the spectrum follows the trend of pure LC. Peak 827~cm$^{-1}$ present in the spectrum of the MgO doped LC system corresponds to the Mg--O--Mg bond vibrations. The peak at 501~cm$^{-1}$ corresponds to Zn--O vibration in the spectrum of ZnO doped~LC~system.

\subsection{XRD spectroscopy}
\begin{figure}[htb]
	\centerline{\includegraphics[scale=0.6]{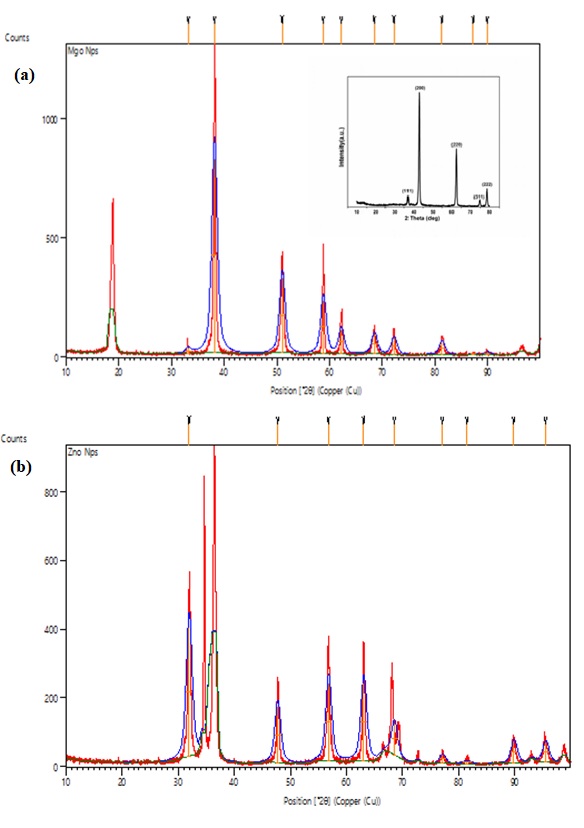}}
	\caption{(Colour online) XRD spectrum of (a) MgO QDs [Inset figure shows standard JCPDS (card no.~89-7746) file for MgO QDs] and (b) ZnO QDs.} \label{fig-smp7}
\end{figure}

We have obtained high resolution X-ray diffraction (HR-XRD) data using Rigaku SmartLab 9kW Powder type (without $\chi$ cradle) X-machine. We have compared the XRD data with standard JCPDS (card no. 89-7746) file, which confirms the formation of a cubic structure. The XRD data confirm the highly crystalline nature of the sample. The polycrystalline cubic phase of MgO nanoparticles is confirmed by the broadened peaks detected at 38.1312°, 50.9804°, 58.793°, 72.196°, and 81.2994° and can be referred to the (111), (200), (220), (311), and (222) crystal planes, respectively. We use Debye-Scherrer formula to compute the size of synthesized MgO QDs given by equation~\eqref{eq_3_1}.
\begin{equation}
	D = \frac{0.94 \lambda}{\beta\cos\theta}.
	\label{eq_3_1}
\end{equation}
Here, $D$ is average crystalline domain size (i.e., average size of NPs), $\lambda$ is the wavelength of the used X-ray, $\beta$ is full width at half maximum (FWHM), and $\theta$ is the diffraction angle. To extract out a larger size MgO NPs, we further filtered NPs by making MgO solution and passed it through mesh of ultra-fine filter via nylon 25~mm 0.2~\textmu{}m (AXIVA). From the data in table~\ref{tbl-smp1}, the smallest size of QDs is 7.0603~nm. The largest size of synthesized QDs is 9.5647~nm. The obtained size range of synthesized MgO QDs via sol-gel method is also comparable to the data observed in reference~\cite{RS31}. Figure~\ref{fig-smp7}(a) depicts the powder XRD pattern of MgO QDs obtained after calcination at 550°C.

 The powder XRD patterns obtained in figure~\ref{fig-smp7}(b) confirm the crystalline structure of ZnO QDs and also confirm its purity. The XRD data are presented in the table~\ref{tbl-smp2}. The observed intensity peaks and their positions closely match that of JCPDS file no. 00-036-1451. The peaks corresponding to different angular positions and different ($hkl$) values confirm the wurtzite hexagonal phase of zinc oxide (ZnO) QDs. Prominent peaks detected at 31.9263°, 47.7099°, 56.7854°, 63.0500°, 68.5059°, 77.0961°, 81.4671°, and 89.8232°can be referred to the planes having Miller indices (100), (101), (102), (110), (103), (200), (112), and (201), respectively. The broadening of the peaks in the XRD spectrum is accounted on the basis of the formation of nanoparticles. No impurity peaks are visible in our XRD data. The average particle size of synthesized nanoparticles lies in the range 7.04 nm to 11.31 nm. The prominent size of QDs is 7.04 nm, having the highest relative intensity and larger $d$-spacings.

\begin{table}[htb]
	\caption{XRD data of synthesized MgO QDs.}
	\label{tbl-smp1}
%	\vspace{2ex}
	\begin{center}
		\renewcommand{\arraystretch}{0}
		% 		\begin{landscape}
			\scriptsize{
				\begin{tabular}{|c|c|c|c|c|c|c|}
					\hline
%S.No.& Position($^{o}2\theta$) & Height(cts) & FWHM($2\theta$(radian)) & Left d spacings($\AA$) & Relative Intensity (\%) & Size of QDs (nm)\strut\\
S.No&Position($2\theta$)&Height(cts)&FWHM(\verb|radian|)($2\theta$)&Left d spacings (\AA)&Relative intensity (\%)&Size of QDs (nm)\\
					\hline
					1& 33.1363 &19.09& 0.0214 &  2.7036 & 2.35 & 7.0603\strut\\
					\cline{1-7}
					2& 38.1312 &810.97& 0.0214 &  2.3601 & 100.00 & 7.1599\strut\\
					\cline{1-7}
					3& 50.9804 &312.99& 0.0214 &  1.7914 & 38.60 & 7.4969\strut\\
					\cline{1-7}
					4& 58.7930 &223.49& 0.0214 &  1.5706 & 27.56 & 7.7672\strut\\
					\cline{1-7}
					5& 62.1978 &99.85& 0.0214 &  1.4926 & 12.31 & 7.9030\strut\\
					\cline{1-7}
					6& 68.4735 &81.75& 0.0214 &  1.3703 & 10.08 & 8.1855\strut\\
					\cline{1-7}
					7& 72.1960 &68.56& 0.0214 &  1.3085 & 8.45 & 8.3750\strut\\
					\cline{1-7}
					8& 81.2994 &53.70& 0.0214 &  1.1835 & 6.62 & 8.9193\strut\\
					\cline{1-7}
					9& 87.2625 &4.57& 0.0214 &  1.1173 & 0.56 & 9.3495\strut\\
					\cline{1-7}
					10& 89.9341 &11.11& 0.0214 &  1.0909 & 1.37 & 9.5647\strut\\
					\cline{1-7}
					\hline
				\end{tabular}
			}
			% \end{landscape}
		\renewcommand{\arraystretch}{1}
	\end{center}
\end{table}

\begin{table}[htb]
	\caption{XRD data of synthesized ZnO QDs.}
	\label{tbl-smp2}
%	\vspace{2ex}
	\begin{center}
		\renewcommand{\arraystretch}{0}
		\scriptsize{
			\begin{tabular}{|c|c|c|c|c|c|c|}
				\hline
S.No & Position($2\theta$) & Height(cts) & FWHM(\verb|radian|)($2\theta$) & Left d spacings (\AA) & Relative intensity (\% )& Size of QDs (nm)\\
%&&&&&&\\
				\hline
				\rule{0pt}{2pt}&&&&&&\\
				\hline
				1& 31.9263 &375.38& 0.0214 &  2.8032 & 100.00 & 7.0385\strut\\
				\cline{1-7}
				2& 47.7099 &162.70& 0.0214 &  1.9063 & 43.34 & 7.3992\strut\\
				\cline{1-7}
				3&56.7854 &226.52& 0.0214 &  1.6213 & 60.34 & 7.6925\strut\\
				\cline{1-7}
				4& 63.0500 &224.61& 0.0214 &  1.4744 & 59.84 & 7.9388\strut\\
				\cline{1-7}
				5& 68.5059 &91.84& 0.0214 &  1.3697 & 24.47 & 8.1871\strut\\
				\cline{1-7}
				6& 77.0961 &25.30& 0.0214 &  1.2371 & 6.74 & 8.6527\strut\\
				\cline{1-7}
				7& 81.4671 &9.14& 0.0286 & 1.1814 & 2.43 & 6.6823\strut\\
				\cline{1-7}
				8& 89.8232 &63.36& 0.0214 &  1.0920 & 16.88 & 11.3088\strut\\
				\cline{1-7}
				9& 95.4681 &57.85& 0.0214 &  1.0418 & 15.41 & 10.0615\strut\\
				\cline{1-7}
				\hline
			\end{tabular}
		}
		\renewcommand{\arraystretch}{1}
	\end{center}
\end{table}

The dominant sizes of synthesized MgO and ZnO QDs are 7.16 nm and 7.04 nm, respectively. When nanoparticles are added to a liquid crystal matrix like 6CHBT, their size can play a significant role. Here, when we discuss individual particle scenario, we prefer to use QDs. However, QDs self-assemble and form larger spherical particles, we refer to them as spherical nanodots. NPs is used for a general theoretical description of the principle involved as it is valid for all NPs.

\subsection{UV–vis–NIR spectroscopy}
\begin{figure}[ht!]
	\centerline{\includegraphics[scale=0.55]{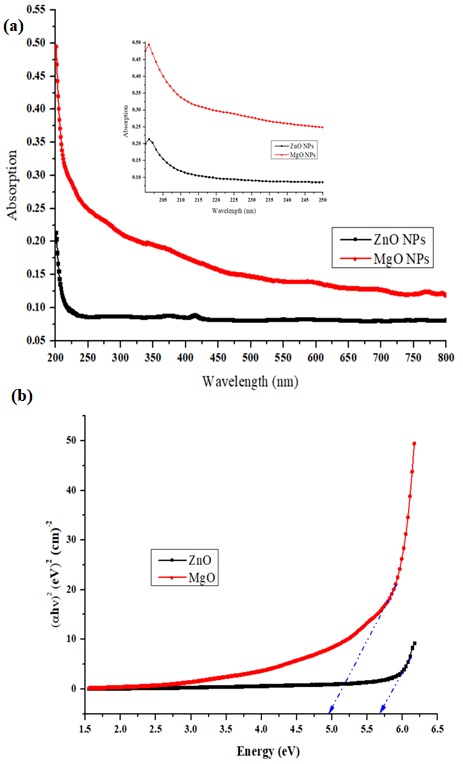}}
	\caption{(Colour online) (a) UV–vis–NIR spectra of synthesized MgO nanoparticle and (b) Tauc’s plot of MgO nanoparticle.} \label{fig-smp8}
\end{figure}

UV–vis–NIR spectra for MgO and ZnO QDs are observed using Shimadazu UV-visible 1900i. In order to measure the absorbance of MgO QDs, UV-visible spectrophotometer is used to record UV-visible spectra of the generated MgO and ZnO nanopowder in the absorbance mode in the wavelength range 200--800~nm as shown in figure~\ref{fig-smp8}(a). The observations were performed by making dissolving QDs in ethanol solution. The absorption peaks are presented in the inset. Broad peak at 202 nm shows that the nanoparticles possess a quantum confinement. We determined the energy band gap of magnesium oxide, using Tauc's relation as in equation~\eqref{eq_3_2}.
\begin{equation}
	%%	alpha hnu = A(hnu-E_{g})^{n}
\alpha h \nu = A(h\nu-E_{g})^{n}.
\label{eq_3_2}
\end{equation}
Here, $n$ is a constant and is equal to $\frac{1}{2}$ for direct bandgap semiconductors, $h$ is the Planck's constant, $A$ is a constant, $E_{g}$ is the energy bandgap, and $\alpha$ is the absorption coefficient. Extrapolating the curve drawn between $h\nu$ and $(\alpha h\nu)^{2}$, the band gaps of the MgO and ZnO QDs were calculated from figure~\ref{fig-smp8}(b). The optical absorption coefficient is indicated by the symbol $\alpha$, and $\nu$ indicates the frequency. The energy bandgaps were also obtained by extrapolating the curve with the help of three tangent lines, and were found to be 4.95 eV and 5.67 eV, respectively for MgO and ZnO QDs. The particle size and bandgap of MgO QD confirm its optoelectronic applications. The obtained bandgap value of ZnO QD is larger than that of MgO QD.

\begin{figure}[htb]
	\centerline{\includegraphics[scale=0.55]{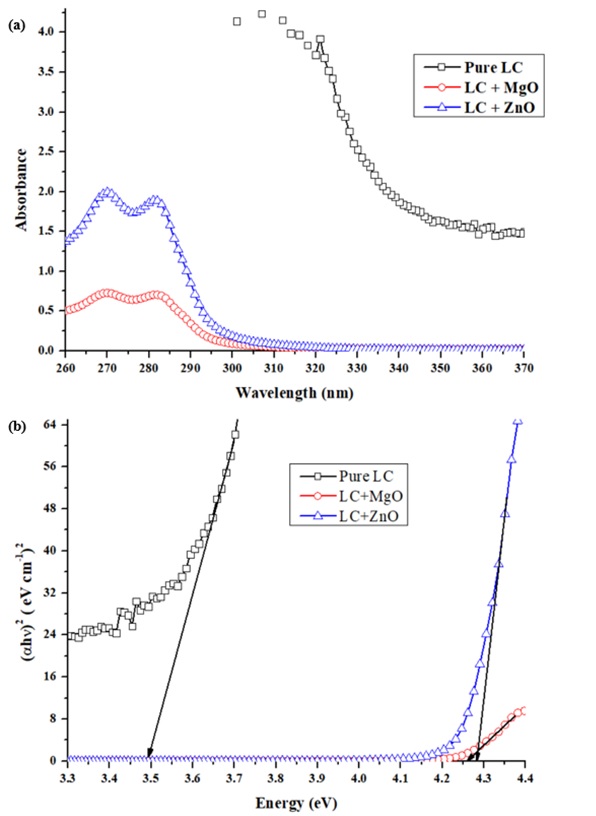}}
	\caption{(Colour online) (a) UV–vis–NIR spectra of pure LC, MgO, and ZnO doped LC system and (b) Tauc’s plot of pure LC, MgO, and ZnO doped LC system.} 
	\label{fig-smp9}
\end{figure}

UV–vis–NIR spectra of pure LC, MgO and ZnO doped LC system is shown in figure~\ref{fig-smp9}(a). After doping NPs, the UV absorption peak of pure LC shifts towards ultraviolet range. We use Tauc plot method for calculation of bandgap of the composite system as shown in figure~\ref{fig-smp9}(b). The bandgap of pure LC is found to be 3.49 eV. After doping MgO and ZnO NPs in LC media, it increases to 4.26 eV and 4.28~eV, respectively. After doping with MgO and ZnO NPs, the bandgap of LCs increases due to a process known as quantum confinement. Quantum confinement occurs when the size of a substance is decreased to nanoscale levels, causing electrons and holes to be confined inside a small space. This confinement leads to distinct energy levels, which raises the bandgap. Depending on their size, the nanoparticles act as quantum wells or quantum dots. These nanostructures limit the charge carriers (electrons and holes) within their boundaries, resulting in discrete rather than continuous energy levels. As a result, the bandgap of the doped liquid crystal system increases. Zn is present in ZnO, and Zn has a higher electronegativity than Mg in MgO. A bigger shift in the energy levels and a wider bandgap are caused by a higher electronegativity of Zn, which interacts more strongly with the nearby atoms in the crystal lattice. ZnO doping that caused a bigger shift of 0.79 eV than MgO doping that caused a shift of 0.77 eV.

\subsection{Optical microscopy}
\begin{figure}[h!]
 	\centerline{\includegraphics[scale=0.6]{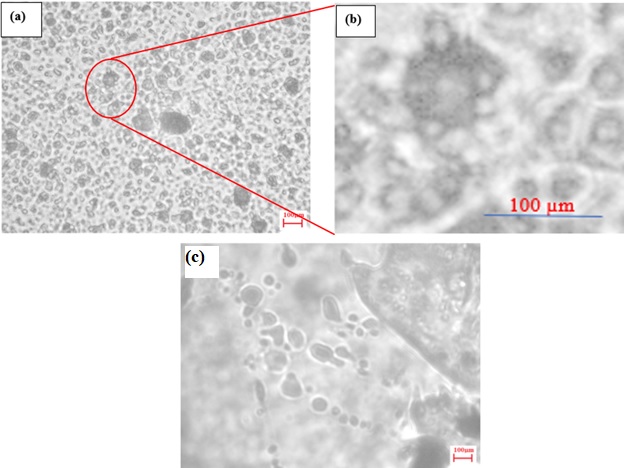}}
 	\caption{(Colour online) Optical microscopy images of the different part of the sample (a) MgO doped 6CHBT LC system, (b) enlarged image of the sample, showing agglomeration of small size MgO NPs (dark) surrounded by anchored LC molecules (bright color)  and (c) showing large size MgO NP formation of aggregates/clusters on glass surface.} \label{fig-smp10}
 \end{figure}

Optical microscope images were obtained using inverted optical microscope (Zeiss) of 0.1~\textmu{}m resolution and objestive lens of 100x oil immersion\verb|/|1.25 and 20x\verb|/|0.30. Optical microscopy images show the agglomeration of MgO NPs and anchoring of LC molecules over its surface. The MgO dot gives dark shadows and dense anchoring of LCs over it gives bright reflections. Small MgO NPs after aggregation form nanodot wire-like structures. These nanodot-wire or one-dimensional arrays of MgO NPs offer their surfaces to LC molecules for anchoring. LC molecules anchoring over NP surface provide a means of linking them as well provide stability to nanodot wire-like structures. The integration of dots into nanodot wire-like nanostructures results in a further interconnected macroscopic network similar to the case of inorganic polymer molecules figure~\ref{fig-smp10}.

\begin{figure}[htb]
	\centerline{\includegraphics[scale=0.65]{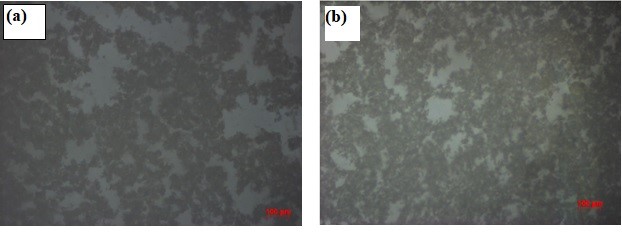}}
	\caption{(a) and (b) Show images of different part of the MgO NPs doped 6CHBT LC sample on the ITO coated glass surface obtained using optical microscope.} \label{fig-smp11}
\end{figure}

 Figure~\ref{fig-smp10} shows the formation of microstructure by self-assembly of NPs in presence of LC molecules on the plane glass surface. Figures~\ref{fig-smp10}(a) and (b) show honeycomb-like microstructures. MgO NPs agglomerate and form large spherical nano-micro-clusters as shown in figure~\ref{fig-smp10}(c). A careful observation of figure~\ref{fig-smp10}(c) reveals that ring-like structures of sizes $\sim $30--40~\textmu{}m are also formed as the agglomerated MgO NPs embedded in host LC media complete the periphery of a complete circle. The spherical nano-micro-clusters arrange themselves to form a circle. The nano-micro-dots are linked and stabilized by the LC molecules shown with brighter color (white shade). The agglomeration of NPs and anchoring of LC molecules on ITO coated glass surface can be clearly observed in figure~\ref{fig-smp11}(a) and figure~\ref{fig-smp11}(b).
  
\begin{figure}[htb]
	\centerline{\includegraphics[scale=0.6]{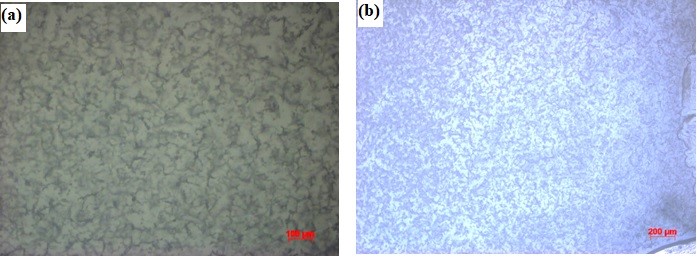}}
	\caption{(Colour online) Optical microscopy images of ZnO doped LC system on (a) plane glass and (b) ITO coated glass surfaces.} \label{fig-smp12}
\end{figure}

Figure~\ref{fig-smp11}(a) and (b) show the effect of ITO coating on the glass surface. It shows interconnected MgO NPs clusters with minimum anchoring of LC molecules at NPs surface. The figure indicates that the ITO attraction to LC molecules is energetically more preferable than that of MgO NPs intermolecular interaction with LC molecules as well as ITO surface. Therefore, the agglomeration of MgO NPs takes place in order to reduce the excess surface free energy of NPs. Mg--O bond strength is high, i.e., 363.2~kJ$\cdot$mol$^{-1}$, and for Mg--Mg is low, i.e., 8.55~kJ$\cdot$mol$^{-1}$ whereas, for Zn--O and Zn--Zn values are 159.0~kJ$\cdot$mol$^{-1}$ and 29.0~kJ$\cdot$mol$^{-1}$, respectively~\cite{RS32,RS33,RS34}. ZnO nanoparticles did not show agglomeration in LC media. Figure~\ref{fig-smp12}(a) shows the optical microscopy image of ZnO NPs doped LC system on glass surface. No hierarchical arrangement of nanoparticles is obtained for the case of ZnO NPs. Figure~\ref{fig-smp12}(b) shows the ZnO NPs doped LC system on ITO coated glass surface. No variations are observed for the ITO coated system as well. In case of ZnO NPs, the intermolecular interaction appears to be weaker than the intramolecular interactions. Therefore, ZnO NPs do not form microstructures by anchoring, and linking with LC molecules rather shows a tendency to form clusters of ZnO NPs. Their long-range aggregation matches more like a percolated assembly. The intermolecular interactions among ZnO NPs are of non-bonding van der Waal force which are far too weak to influence NP aggregation, at least non-volatile or semi-volatile conditions. The dipole-dipole interactions in ZnO NPs are low. In comparison to the H-bond, the intermolecular forces are weak. Important thing is to note here that very little quantitative knowledge is available about the interaction of LC molecules with metal atoms and their oxide NPs. Their anchoring properties of metal oxide NPs are therefore not well understood, though we have tried to develop a qualitative basis here to account the reasons for the existence of such pattern variations. 
 
\subsection{Raman and IR spectroscopy}
\begin{figure}[htb]
	\centerline{\includegraphics[scale=0.4]{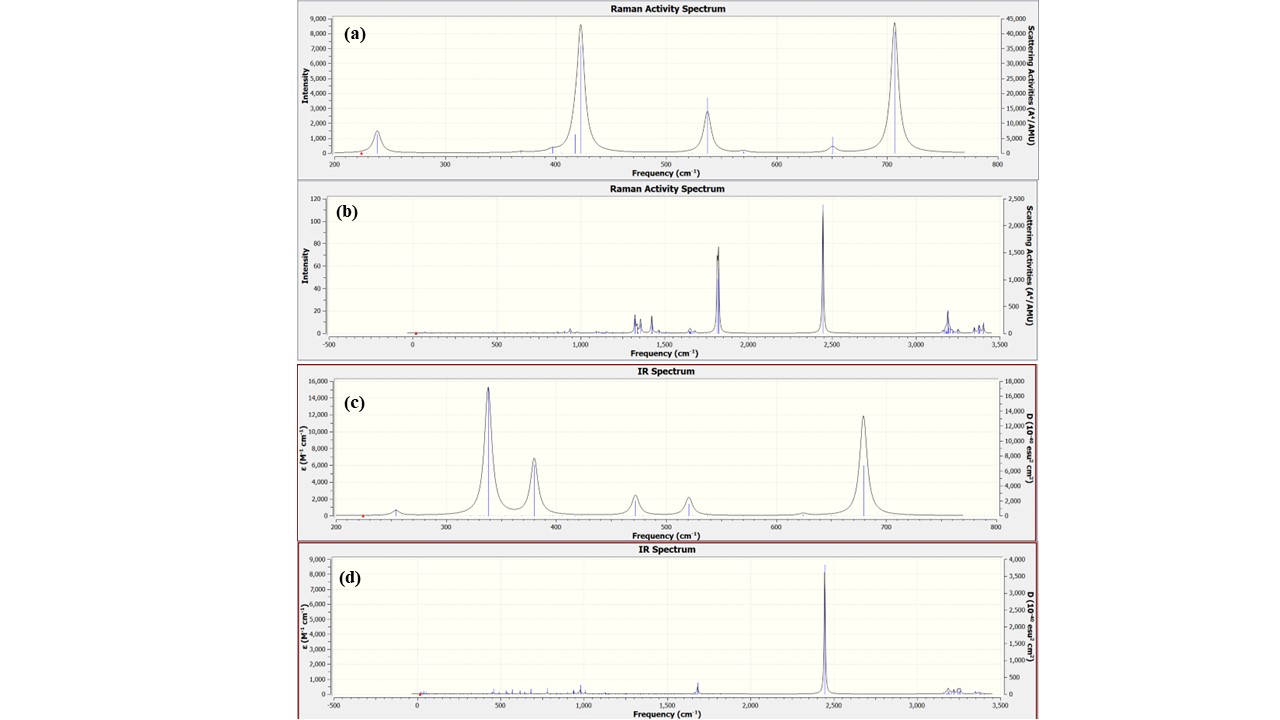}}
	\caption{(Colour online) IR spectra of (a) pure MgO nanoparticle and (b) pure 6CHBT liquid crystal. Raman spectra of (c) pure MgO nanoparticle and (d) pure 6CHBT liquid crystal.} \label{fig-smp13}
\end{figure}

\begin{figure}[htb]
	\centerline{\includegraphics[scale=0.8]{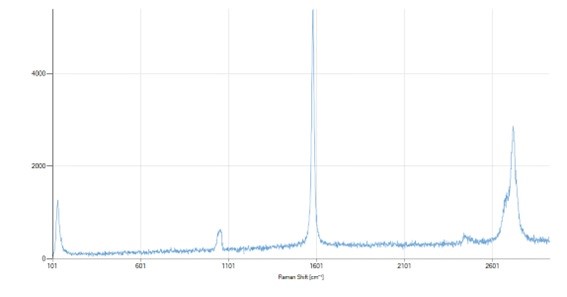}}
	\caption{(Colour online) Raman spectra of MgO doped liquid crystal hybrid system obtained experimentally.} \label{fig-smp14}
\end{figure}

IR spectrum of MgO nanoparticle was obtained theoretically as shown in figure~\ref{fig-smp13}(a). It is obtained using Gaussian software. We have used HF method and 6--31G basis set for this calculation. It shows IR active in the frequency range 250~cm$^{-1}$ to 700~cm$^{-1}$. IR spectrum of pure liquid crystal is shown in figure~\ref{fig-smp13}(b). The prominent peak is obtained at 2445~cm$^{-1}$ corresponding to -N=C=S stretching vibration. The Raman activity is also calculated using the same method and the obtained spectrum is shown in figure~\ref{fig-smp13}(c). The MgO crystal is Raman active in the frequency range 230~cm$^{-1}$ to 750~cm$^{-1}$. The prominent peak due to MgO is observed nearly at 710~cm$^{-1}$. Raman spectrum of pure LC molecule is observed as shown in figure~\ref{fig-smp13}(d). The experimental observation of prominent peaks of the MgO doped 6CHBT liquid crystal is shown in figure~\ref{fig-smp14}. The spectrum is gathered in the range 101~cm$^{-1}$ to 2901~cm$^{-1}$. The -N=C=S stretching vibrations of LC molecules are accountable for the peak at 1172~cm$^{-1}$. C--C and C=C aromatic bond stretching is responsible for the peaks between 650~cm$^{-1}$ and 1650~cm$^{-1}$. Small absorption peak at 1502~cm$^{-1}$ confirms the presence of benzene ring of the LC molecule. C=C stretching vibrations are associated with peaks in the 1600~cm$^{-1}$ to 1660~cm$^{-1}$ range. The peak at 1600~cm$^{-1}$ verifies the presence of C=C stretching vibrations. The appearance of diffusive peaks near 2400~cm$^{-1}$ indicates that the LC molecule behaves like a liquid.

Although the theoretical observations are obtained in gaseous medium, the experimental data are obtained in ethanol medium. We compared the Raman spectra of pure MgO NPs and pure LCs with the experimentally observed Raman spectra of the hybrid system. The Gaussian results obtained in a gas phase are valid for isolated molecules. Therefore, the position of peaks may significantly differ from the measured experimental data and one cannot quantitatively determine the effect of interactions between LC and NPs in hybrid materials, though a qualitative understanding may appear alright. However, the comparisons reveal that the presence of MgO bond stretching in LC medium is represented by peaks at 150~cm$^{-1}$ and 1090~cm$^{-1}$. The intense theoretical MgO peak is observed at nearly 710 cm$^{-1}$ which corresponds to the experimentally observed peak at 1090~cm$^{-1}$. The presence of CH$_{2}$ and CH$_{3}$ in the LC molecule is confirmed by the presence of a peak near 2800~cm$^{-1}$. The modest strong peaks between 1200~cm$^{-1}$ and 1602~cm$^{-1}$ in figure~\ref{fig-smp14} show LC molecule anchoring to the nanodot surface, resulting in a relatively high density of LC molecules locally at the MgO QDs platelets or in their vicinity.

\subsection{FESEM analysis}
\begin{figure}[htb]
	\centerline{\includegraphics[scale=0.7]{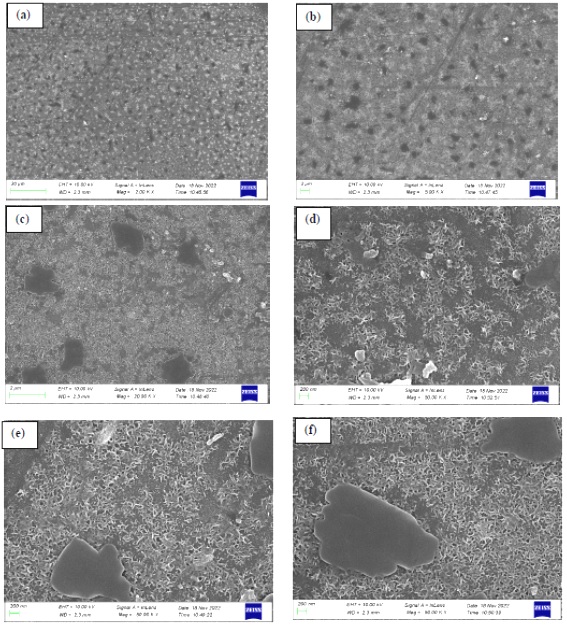}}
	\caption{Field emission scanning electron microscopy (FESEM) images of the different part of the MgO NPs doped LC sample in different magnifications.} \label{fig-smp15}
\end{figure}

FESEM data were obtained using SUPRA 40 VP (Zeiss) equipment.The surface morphologies of MgO doped 6CHBT LC system on ITO coated glass surface can be observed clearly by FESEM images. MgO nanoparticles agglomerate and LC molecules stick on the NPs surface and form flower-like or rice grains-like morphology. Figure~\ref{fig-smp15} shows different part of  the samples in different resolutions. MgO NPs shows agglomeration and forms micro-platelets-like morphology. The size of these platelets-like morpholgy are in the range 100 nm up to 2~\textmu{}m. The platelets also resemble the flower-like structures. Very small sized NPs show agglomeration forming the platelets-like morphology. In lower resolutions, figure~\ref{fig-smp15}(a), \ref{fig-smp15}(b) and \ref{fig-smp15}(c), islands of regions formed by LC molecules (dark patch) and surrounded by dispersed NP rich regions can be observed. In higher resolutions, figure~\ref{fig-smp15}(d), \ref{fig-smp15}(e), and \ref{fig-smp15}(f), the fixed nano-size elongated grain-like aggregation of MgO nanoparticles can be clearly observed. 

One may observe a diffuse macroscopic picture of LC rich regions separated by MgO rich region indicating a macroscopic size reorganization and restructuring. Grains of MgO NPs are interconnected like polymeric chains as illustrated in the figure~\ref{fig-smp2}. Rememeber that a hometropic and homgenous structure formation over the surface of a NP may also exist, and not eseentialy on planar surfaces. The observed structures suggest a homeotropic type of growth at NP surface to a great extent. This can be explained using figure~\ref{fig-smp2}.  In figure~\ref{fig-smp15}(c) and \ref{fig-smp15}(d), one may observe occasional agglomeration of MgO NPs forming large clusters of sizes 200--300~nm or even extending up to a few microns. Since, the smaller nanoparticles do not disturb LCs equilibrium to a great extent, the smaller MgO NPs develop a synergy with LC equilibrium, and the formation of inorganic polymer chain-like network permits their accomodation in LC media. A reconciliation between the driving forces, as discussed in detail in the introduction section, is achieved. The larger size nanoparticles cannot compromise or maintain the energy balance with the LC molecules, and are thus forced out and nucleate to form bigger size NPs aggregate as in figure~\ref{fig-smp15}(e) and \ref{fig-smp15}(f). These LC molecules are attached at the boundaries of the surface of the nanoclusters.

Due to strong dipolar ineteraction between LC molecules and MgO NPs, LC molecules reorient themselves and due to an attractive intramolecular interaction, LC molecules stick to the NPs suraface. The specific surface area (i.e, surface to volume ratio) of the nanomaterials increases drastically when their sizes reduce, e.g., for cubic crystal it varies as ${6}/{l}$, for spherical ones it varies as  ${3}/{r}$ and for rod-like particles it varies as ${2}/{r'}$.  Here, $l$ is the edge of cubic, $r$ is radius of the spherical, and $r'$ is radius of cylindrical or wire-like nanomaterials. Thus, the surface energy of NPs drastically  increase. The growth or agglomeration of NPs helps to reduce the excess surface energy34. These principles and propetries of NPs also help one to understand the underlying physical phenomena in these hybrid systems.

\begin{figure}[htb]
	\centerline{\includegraphics[width=0.65\textwidth]{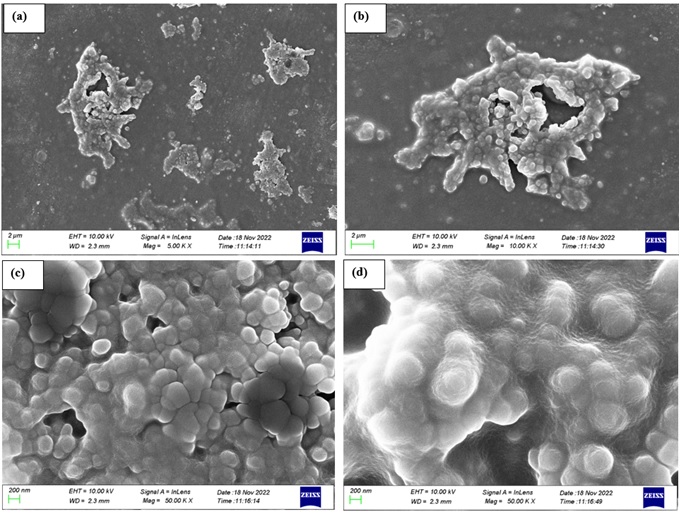}}
	\caption{Field emission scanning electron microscopy (FESEM) images of the different part of the ZnO NPs doped LC sample in different magnifications.} \label{fig-smp16}
\end{figure}

FESEM data of ZnO doped LC system are presented in figure~\ref{fig-smp16}. The observed data are for the film prepared on the ITO coated glass surface. The ZnO nanoparticles agglomerate and form an array of nanospheres extending into irregular shapes as shown in figure~\ref{fig-smp16}(a) and (b). The size of ZnO QDs obtained from XRD data is 7.04 nm. These ZnO QDs agglomerate and form nanospheres or nanodots of diameter $\sim 200$~nm as shown in figure~\ref{fig-smp16}(c) and (d). These nanospheres further agglomerate due to the coupling and form an irregularly arranged array of spheres, figure~\ref{fig-smp16}(c). The anchoring of LC molecules is at the nanoparticles surface, especially at the boundary of the island formed due to the arrangements of nanodots. The presence of LC molecules is clearly observed in figure~\ref{fig-smp16}(d). In case of ZnO NP doping in LC media at ITO surface, we do not observe any fine and unique arrangements of nanodots/nanospheres formed with them. This can be accounted on the basis of the interparticle interactions and interfacial energy contributions. In case of ZnO NPs, the QDs first prefer to form stable nanospheres of dimensions of $\sim 200$~nm due to strong QDs--QDs interactions and then the dispersion forces drive them to form larger island-like agglomeration on ITO surface. This indicates that the LC media cannot host these large size nanospheres, which are energetically not preferable.

\begin{figure}[htb]
	\centerline{\includegraphics[width=0.65\textwidth]{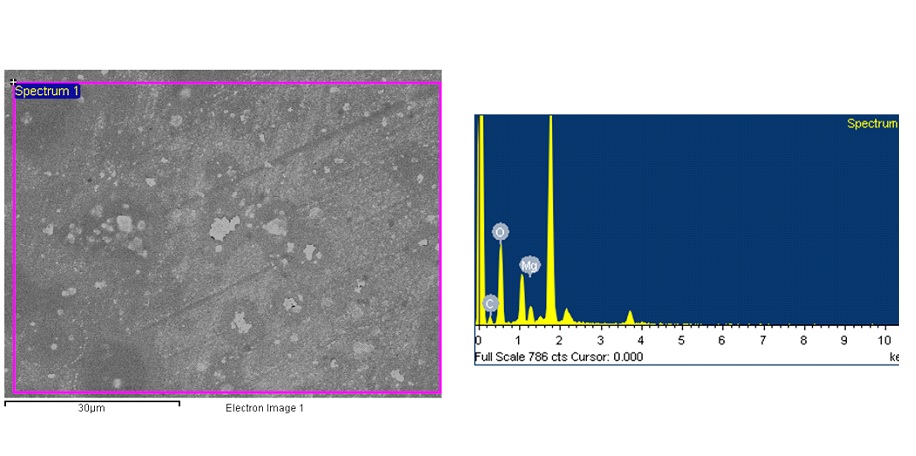}}
	\caption{(Colour online) EDX analysis of MgO doped 6CHBT liquid crystals.} \label{fig-smp17}
\end{figure}

\subsection{Energy dispersive X-ray (EDX) analysis}
Elemental analysis of the sample was confirmed via EDX analysis as shown in figure~\ref{fig-smp17}. Here, we observed the presence of Mg as 10.31 wt\%, O as 75.85 wt\% and C as 13.84 wt\% for K lines as shown in table~\ref{tbl-smp4}. Presence of Mg and O confirms the formation of MgO on the glass surface. EDX data confirm the presence of both LC molecules and MgO. The 6CHBT LC molecule (C19H27NS) shows the elemental weight percentage in energy bands 0.2 to 1.4~keV corresponding to K lines. Here, for LC molecule only a  carbon element is observed in the spectrum. Thin gold coating is done at the films done for EDX analysis. The data were taken by creating vacuum inside FESEM-EDX [SUPRA 40 VP (Zeiss)] using Argon gas.

\begin{figure}[htb]
	\centerline{\includegraphics[width=0.65\textwidth]{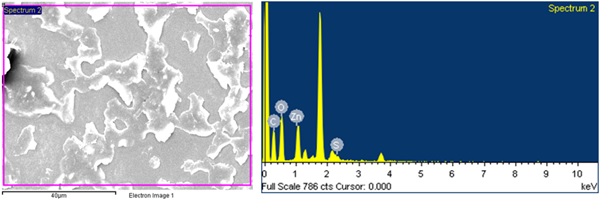}}
	\caption{(Colour online) EDX analysis of ZnO doped 6CHBT liquid crystals.} \label{fig-smp18}
\end{figure}

\begin{table}[htb]
	\caption{EDX data of the MgO NPs doped LC sample.}
	\label{tbl-smp4}
%	\vspace{2ex}
	\begin{center}
		\renewcommand{\arraystretch}{0}
		\begin{tabular}{|c|c|c|}
			\hline
			Element& Weight (\%)&Atomic (\%)\strut\\
			\hline
			\rule{0pt}{2pt}&&\\
			\hline
			C K& 13.84 &18.24\strut\\
			\cline{1-3}
			O K& 75.85 &75.05\strut\\
			\cline{1-3}
			Mg K& 10.31 &6.71\strut\\
			\cline{1-3}
			Totals& 100.0 &100.0\strut\\
			\cline{1-3}
			\hline
		\end{tabular}
		\renewcommand{\arraystretch}{1}
	\end{center}
\end{table}

\begin{table}[htb]
	\caption{EDX data of the ZnO NPs doped LC sample.}
	\label{tbl-smp5}
	%\vspace{2ex}
	\begin{center}
		\renewcommand{\arraystretch}{0}
		\begin{tabular}{|c|c|c|}
			\hline
			Element& Weight(\%)&Atomic (\%)\strut\\
			\hline
			\rule{0pt}{2pt}&&\\
			\hline
			C K& 37.26 &50.11\strut\\
			\cline{1-3}
			O K& 44.93 &45.37\strut\\
			\cline{1-3}
			S K& 0.45 &0.22\strut\\
			\cline{1-3}
	    	Zn K& 17.36 &4.29\strut\\
			\cline{1-3}
			Totals& 100.0 &100.0\strut\\
			\cline{1-3}
			\hline
		\end{tabular}
		\renewcommand{\arraystretch}{1}
	\end{center}
\end{table}

Elemental analysis of ZnO doped LC system was confirmed via EDX analysis as shown in figure~\ref{fig-smp18}. Here, we observe the presence of Zn in 17.36 wt\%, O in 44.93 wt\%, S in 0.45 wt\% and C in 37.26 wt\% for  K lines of X-ray used in EDX data. The data are presented in table~\ref{tbl-smp5}. Presence of Zn and O confirms the formation of ZnO on the glass surface. EDX data confirm the presence of both LC molecules and ZnO.

\subsection{Differential scanning calorimetric measurement for phase transition }
\begin{figure}[ht!]
	\centerline{\includegraphics[scale=0.55]{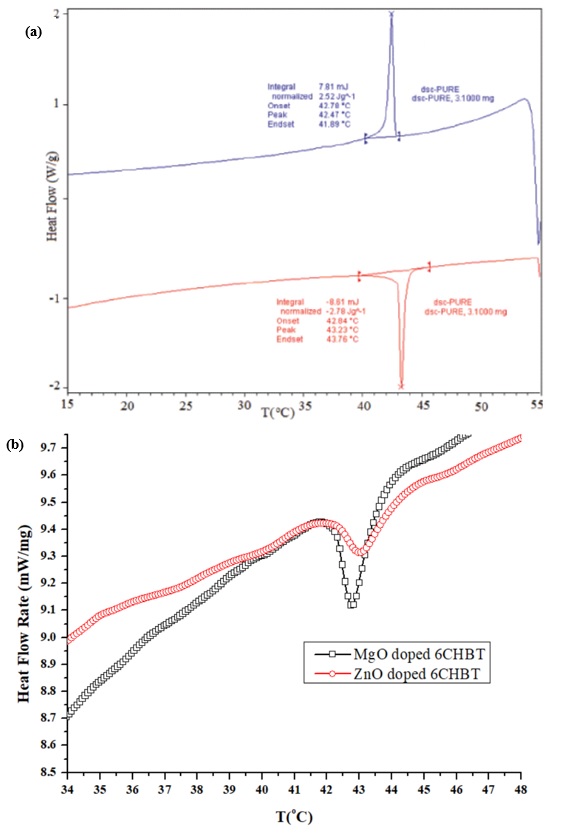}}
	\caption{(Colour online) DSC thermogram of (a) pure 6CHBT liquid crystals representing nematic to isotropic phase transition in heating and cooling mode. %Reproduced with permission from reference~[?]. 
		(b) Calculated DSC thermograms for MgO and ZnO NPs doped LC system representing nematic to isotropic phase transition in the heating mode.} \label{fig-smp19}
\end{figure}

A typical DSC thermogram for pure LCs is presented in figure~\ref{fig-smp19}(a) (reproduced from reference~\cite{RS13}). The clearing point temperature  of pure 6CHBT is about 43.23°C. We have obtained the DSC data for MgO and ZnO doped LC system, using DSC-60 Plus machine. We have obtained the DSC thermogram for doping concentrations of 0.5. The DSC thermogram MgO and 0.5. The DSC thermogram ZnO nanoparticles in the LC molecules, figure~\ref{fig-smp19}(b). The incorporation of MgO and ZnO nanoparticles into the liquid crystal (6CHBT) matrix influences the transition temperatures between the nematic and isotropic phases. The transition temperatures for nematic to isotropic transition for MgO and ZnO nanoparticles doped samples are found to be 42.82°C and 43.09°C, respectively. The nematic isotropic transition for pure 6CHBT LC occurs at 43.23°C. Nanoparticles alter the molecular ordering and phase behavior of liquid crystals. Nanoparticles serve as nucleation sites and interfere with the ordering of liquid crystal molecules, influencing the phase transition temperatures. In our case, the addition of nanoparticle results in a slightly lower transition temperature from the nematic to the isotropic phase. This behavior is related to the heterogeneous nucleation effect, in which nanoparticles act as nucleation sites and assist the transition. The presence of nanoparticles lowers the energy barrier for the transition, resulting in a lower transition temperature as compared to a pure liquid crystal. The incorporation of MgO nanoparticle lowers the transition point more than that of ZnO nanoparticles.

\subsection{Dielectric permittivity of doped LC system }
Liquid crystals possess both liquid and crystalline behaviour. They are anisotropic, which means that their physical qualities change depending on the direction. Liquid crystal dielectric characteristics are also anisotropic and are affected by factors such as temperature, frequency, and the molecular structure of the liquid crystal compound. One of the most essential dielectric qualities is the dielectric constant, commonly known as relative permittivity. It describes a material's capability to store electrical energy in an electric field. The dielectric constant of a liquid crystal varies depending on its alignment and phase (e.g., nematic, smectic, cholesteric). The presence of nanoparticles causes changes in the distribution of the local electric field inside the liquid crystal matrix. This can result in a changed dielectric constant, which affects the material's capacity to store and transport electrical charges. Depending on their surface qualities, nanoparticles behave as conductive or insulating zones inside the matrix. This has an effect on the material's charge storage and discharge characteristics, impacting its capacitive behaviour. The electrical conductivity of the 6CHBT matrix is influenced by MgO and ZnO nanoparticles. The presence of nanoparticles help or hinder the charge transfer within a composite material. The interaction of nanoparticles and liquid crystal molecules alters the charge carrier mobility, resulting in altered electrical conductivity.

\begin{figure}[ht!]
	\centerline{\includegraphics[scale=0.55]{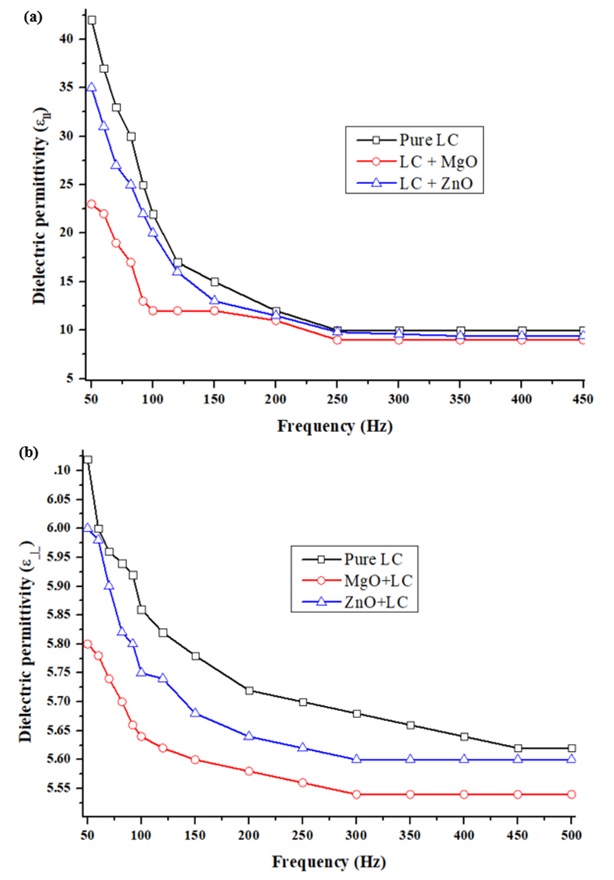}}
	\caption{(Colour online) Variation of (a) parallel, and (b) perpendicular components of dielectric permittivity of pure 6CHBT LC, ZnO, and MgO doped LC system at 32°C.} \label{fig-smp20}
\end{figure}

Figure~\ref{fig-smp20} displays the dielectric permittivity component curves for pure MgO and ZnO doped 6CHBT at 32°C, respectively. The collected trend is comparable with the previously obtained data in reference~\cite{RS13}. We performed the experiment using Impedance analyzer model Alpha-A-High Performance Frequency Analyzer and observed the data. With the increase of frequency in the 50--200~Hz range, the dielectric permittivity of both pure and doped NLC systems decreases significantly. Pure 6CHBT possesses high dielectric permittivity in the low-frequency region, which is further decreased by doping ZnO and MgO nanoparticles. The dielectric permittivity of liquid crystals is determined by the induced dipole moment and the polarization caused by orientation of molecules. Ion dispersion is homogeneous in liquid crystals as well. The frequency dependence of the nonuniform ion distribution caused by the application of the electric field also affects the dielectric permittivity. Additionally, it has been found that the bulk and the region near the surface have higher space charge densities at lower frequencies. As a result, the impact of space charge is greater at lower frequencies. The addition of MgO NPs reduces the dielectric permittivity of the pure LC system in comparison with that of ZnO NPs.

\section{Conclusion}
We synthesized the fine and small MgO and ZnO QDs via sol-gel method. We confirmed the purity, crystalline nature apart of particle sizes of MgO and ZnO nanoparticles using FTIR, Raman, and IR spectroscopy. The elemental analysis was done using FESEM-EDX which also confirmed the formation of MgO and ZnO doped LC hybrid systems. The SEM analysis confirms the associated finer morphologies of the microstructures due to self-assembly of NPs in LC media. The structures observed can be classified into three main categories: grain-like structures formed by aggregation of smaller size MgO NPs while LC molecules anchor over the NPs surfaces; the grain-like structures further integrate to form inorganic polymeric type of honeycomb-like mesostructures in presence of glass surface, and flower-like clusters of MgO NP clusters on ITO surface. It was found that the hierarchical arrangement of MgO NPs in presence of glass and ITO surfaces is completely different. The ZnO NPs agglomerate in another fashion and have no similarties with MgO NPs self-assembly. We account reasons for the differences in their hierarchical structures on the basis of competetive molecular forces. There appears a competing effect of the forces, the one which lowers the energy of the hybrid system wins the game. The ITO surface appears to have a greater affinity to LC than glass, and, thus the anchoring of LC molecules on MgO NPs is reduced in that case. MgO NPs are not linked via LC molecules and then try to form clusters to minimize their excess surface free energies. We further see a self-assembly as an important tool for the synthesis of composite nanomaterials and designing and integrating them to make novel devices. The novel microstructures having unique functional properties may find vital applications in future. We have discussed and explored the underlying science behind the phenomenological differences in the case of MgO and ZnO NPs on glass and ITO coated glass surfaces.
	
\section*{Acknowledgement}

We are thankful to Material Research Centre at MNIT, Jaipur, India for providing us characterization data. Authors are thankful to Centre for Instrumentation (CIF), Banaras Hindu University, Varanasi, India for providing us the FTIR and HR-XRD data. Authors are also thankful to Dr.~Kamlendra~Awasthi, Department of Physics, MNIT, Jaipur, India and Professor A. K. Singh, Department of Physics, Banaras Hindu University, Varanasi, India for extending their support. We are thankful to Mr. Abhilash Bajpai, Nanoscience Centre, IIT Kanpur, India for providing us the FESEM-EDX data.
%\subsection{Conflict of Interest:}
%The authors have no conflict of interest. 
%\subsection{Funding:}
%No funding has been received from any source to carry out the research work.
%\subsection{Reference}

%\bibliographystyle{cmpj}
%\bibliography{cmpjxampl}

%\begin{thebibliography}{86}

%\end{thebibliography}

\newpage

\ukrainianpart

\title{Формування ієрархічних структур нано- та мікромасштабу у рідкокристалічних середовищах, легованих квантовими точками MgO та ZnO: роль конкуруючих сил}

\author{А. К. Сінгх, С. П. Сінгх}
\address{
	Факультет фізики та матеріалознавства, Технологічний університет Мадан Мохан Малавія, Горакхпур-273010, Уттар-Прадеш, Індія
}
%
%% якщо автор є один або автори є з однієї установи:
%
%  \author{1й Автор, 2й Автор, \ldots}
%  \address{Інститут\ldots}
%
%%

\makeukrtitle

\begin{abstract}
	\tolerance=3000%
	У даній статті досліджено вплив легування наночастинками ZnO та MgO 4-(транс-4-н-гексилциклогексил) ізотіоціанатобензоату.
Здійснено ретельне порівняння діелектричних властивостей, а також опто\-електронних та калориметричних властивостей фазового переходу для рідких кристалів, легованих наночастинками MgO та ZnO. 
Приготовлено гомогенну суміш наночастинок MgO та ZnO у толуолі, яку далі перенесено у комірки з простого скла та скла, покритого індієм з оксидом олова (ITO). 
Спостережувані мікро\-структури в гібридній системі можна класифікувати за трьома основними категоріями: зернисті струк\-тури, утворені агрегацією наночастинок MgO менших розмірів, коли молекули рідкого кристалу закріплюються на поверхні наночастинок; grtu-зернисті структури, які далі об'єднуються та утворюють неорганічний полімерний тип пористих структур на скляній поверхні; квіткоподібні кластери наночастинок MgO на поверхні ITO. 
Наночастинки меншого розміру можуть підтримувати енергетичний баланс, дозволяючи молекулам рідкого кристалу закріплюватися на їхніх поверхнях, тоді як наночастинки більшого розміру не можуть порушувати або підтримувати цей баланс з молекулами рідкого кристалу та відокремлюються для зародження й утворення наноагрегатів або кластерів більшого розміру. Перевага поверхні підкладки та наночастинок в енергії відносно молекул рідкого кристалу відіграє важливу роль у формуванні різних типів ієрархічних нано- та мікроструктур.  Розглянуто причини формування ієрархічних структур  нано- та мікромасштабу на основі конкуренції взаємодій ``наночастинка-наночастинка'', ``рідкий кристал--рідкий кристал'', ``наночастинка--рідкий кристал'', ``скло/ITO-наночастинка'' та ``скло/ITO-рідкий кристал''.
Спостерігаються значні зміни в діелектричних властивостях, температурі переходу, ширині забороненої зони та інших параметрах молекул рідких кристалів, легованих наночастинками MgO, в той час як відповідні зміни в рідких кристалах, легованих наночастинками ZnO, є незначними. Дані оптичної мікроскопії, а також експериментів з FTIR, комбінаційного, інфрачервоного, HR-XRD і FESEM-EDX розсіювання підтверджують наші головні висновки.%
	
	\keywords зернисті структури, квіткоподібні мікроструктури, стільникові мезоструктури, наночастинки ZnO, наночастинки MgO
	
\end{abstract}

\lastpage
\end{document}